\documentclass[12pt]{article}
\usepackage{graphicx}
\usepackage{verbatim}

\input epsf
\def\beq{\begin{eqnarray}}
\def\eeq{\end{eqnarray}}

\def\ba{\begin{eqnarray}}
\def\ea{\end{eqnarray}}
\def\beq{\begin{eqnarray}}
\def\eeq{\end{eqnarray}}

\def\bh{\bar{h}}

\def\p{{\cal P}}

\def\L*{{\cal L}_*}
\def\L{\mathcal{L}}
\def\({\left(}
\def\){\right)}

\def\p{\partial}

\def\lsim{\mathrel{\rlap{\lower3pt\hbox{\hskip0pt$\sim$}}
     \raise1pt\hbox{$<$}}}         
\def\gsim{\mathrel{\rlap{\lower4pt\hbox{\hskip1pt$\sim$}}
     \raise1pt\hbox{$>$}}}         

\def\lsim{\mathrel{\rlap{\lower3pt\hbox{\hskip0pt$\sim$}}
     \raise1pt\hbox{$<$}}}         
\def\gsim{\mathrel{\rlap{\lower4pt\hbox{\hskip1pt$\sim$}}
     \raise1pt\hbox{$>$}}}         

\usepackage{amsmath}
\usepackage{amsfonts}
\usepackage{verbatim}

\begin{document}

\begin{center}


{\Large\bf  Classicalization of Gravitons and Goldstones}
 \vspace{0.2cm}

\end{center}

\begin{center}

{\bf Gia Dvali}$^{a,b,c,d}$,  {\bf Cesar Gomez}$^{a,e}$ and 
{\bf Alex Kehagias}$^{f}$

\vspace{.6truecm}

\vspace{.2truecm}

{\em $^a$Arnold Sommerfeld Center for Theoretical Physics\\
Department f\"ur Physik, Ludwig-Maximilians-Universit\"at M\"unchen\\
Theresienstr.~37, 80333 M\"unchen, Germany}


{\em $^b$Max-Planck-Institut f\"ur Physik\\
F\"ohringer Ring 6, 80805 M\"unchen, Germany}


{\em $^c$CERN,
Theory Division\\
1211 Geneva 23, Switzerland}


{\em $^d$CCPP,
Department of Physics, New York University\\
4 Washington Place, New York, NY 10003, USA}


{\em $^e$
Instituto de F\'{\i}sica Te\'orica UAM-CSIC, C-XVI \\
Universidad Aut\'onoma de Madrid,
Cantoblanco, 28049 Madrid, Spain}\\

{\em $^f$
Physics Division, National Technical University of Athens, \\
15780 Zografou Campus, Athens, Greece} \\

\end{center}



 

\centerline{\bf Abstract}

 We establish a  close parallel between classicalization  of gravitons  and 
 derivatively-coupled Nambu-Goldstone-type scalars.   We show, that black hole formation in high energy scattering process 
represents classicalization with the classicalization radius given by 
 Schwarzschild radius of center of mass energy, and with the precursor of black hole entropy 
 being given by number of  soft quanta  composing  this  classical configuration.   
 Such an entropy-equivalent is defined for scalar  classicalons also and is responsible 
 for exponential suppression of their decay into small number of final particles.      
  This parallel works in both ways.  For optimists that are willing to  hypothesize  that 
 gravity may indeed self-unitarize at high energies  via black hole formation, it illustrates that  the Goldstones may not be much  
 different in this respect, and they classicalize essentially by similar dynamics as gravitons.  In the other direction,  
 it may serve as  an useful de-mystifier of via-black-hole-unitarization  process and of the role of entropy in it, as it illustrates, that
  much more prosaic  scalar theories  essentially do the same. 
  Finally, it illustrates that in both cases classicalization is the defining property for 
 unitarization,  and that it sets-in before one can talk about 
 accompanying  properties, such as entropy and thermality of static classicalons (black holes).  
 These properties are by-products of classicalization,  and their equivalents can be defined 
 for non-gravitational cases of classicalization. 


\newpage

\section{Introduction}

   Recently,  a concept of  non-Wilsonian UV-self completeness for a 
  class of non-renormalizable theories   was introduced in \cite{class1, class2, class3}.    The meaning of this concept is,
 that  a seemingly unitary-violating theory   prevents us from going to sub-cutoff distances, by becoming 
classical in deep UV.   This concept was  inspired by the original suggestion  
for Einstein gravity \cite{giacesar}, due to generically-expected 
feature, that high-energy collisions there produce  black holes.   It was  suggested,  that  
this fact leads us to self-completeness since black hole formation automatically prevents us from 
probing sub-Planckian distances.   
 Inspired by this  property of gravity,  in \cite{class1} this concept of self-UV-completeness, which we refer to as 
classicalization,  was generalized to a  class 
of derivatively-self-interacting theories.   Defining property of such theories is energy  
self-sourcing and existence of classical radius that sets the interaction range at high center of mass energy and
 dominates over all the relevant quantum length-scales.  

    Satisfactory understanding of  physical viability of  the classicalization phenomenon requires 
 field-theoretic analysis based on combination of perturbative and non-perturbative methods.  
  Subsequent studies \cite{class2,class3} based on non-perturbative analysis of scattering process, confirm  
the emergence of the classical length scale in such processes.   
In what follows,  we shall also employ  such analysis.   A promising approach in this direction would be to 
develop a path integral formulation  of the problem,  as suggested recently in \cite{goran}.       
   
 The main purpose of the present work is to establish a deeper connection  between 
 classicalization phenomena in gravity and in Goldstone-type scalar theories, for the sake of improving 
understanding on both sides. 
 
  The idea of classicalization, of course, raises number of important question, one of which is, how 
 close are the parallels between gravity and Goldstones? 
 
  In other words there are two major questions. 

$~~~$
  
   1) Do black holes really unitarize scattering amplitudes, and if yes,  why? 
 
 $~~~$  and 
 
   2)  How special are black holes as compared to other  classical configurations 
 for unitarization of the amplitudes? 
 
 $~~~$
 
   These questions are closely related and (as we shall see) the answer to the first question 
is also a key for answering the second one. 
  In order to outline our claims and findings,  let us deconstruct  the above questions.
    
  The  known properties of the black holes are: 1) (semi)classicality; 2) classical stability; 3)
  existence of Schwarzschild (or other)  horizon;  and  4) the subsequent properties:  Absence of classical hair, 
 existence of Hawking temperature and of Bekenstein-Hawking entropy.  
 
     All the properties listed in 4) are the consequences of  1)-3). 
  But obviously,  the property 1) (and 2)) can exist without the rest.  In order to  understand which of these 
 properties are crucial for  unitarization, let us outline why we even expect that black holes can unitarize  
high energy scattering amplitudes.   
  
  The idea is simple.  In order to probe short distances one 
   needs to do high-energy scattering with small impact parameter.   This requires bringing a lot of energy within a small volume. 
 Because any form of energy sources gravity, bringing a lot of energy within the  small volume creates a classical 
gravitational field.  If volume is small (equivalently if energy is large)  so that energy concentration is within 
its Schwarzschild radius $r_g$, the whole region is expected to collapse into a black hole. The  inevitability 
 of the very last step for the 
case of arbitrary  topology  is still under debate, but we shall not enter there. Our findings are precisely
 directed to bypass this issue and  reduce the unitarization process to its bare essentials  
(for which, as we shall argue,  what matters  is not a "blackholeness" but  rather classicality).   
 So therefore, let us pick up 
a spherically-symmetric  geometry in which case most of us probably would agree that black hole formation is inevitable.   

 Now, once black hole is formed, its decay into any few-particle quantum state 
is exponentially suppressed.  Whatever property  is behind this suppression, is the key to the unitarization process.  
 For the  black holes this suppression can be understood from any of the properties 1-4) listed above. 
 And in particular, the suppression can be easily understood from the properties of  entropy and thermality, because 
 probability to produce a pair of highly energetic particles with energy equal to an entire mass of a black hole 
 ($M_{BH}$)  in a thermal evaporation process at temperature $T_H^{-1} \sim M_{BH}L_P^2$ 
 is at least Boltzmann suppressed by a factor
 \begin{equation}
   \Gamma_{BH \rightarrow 2} \, \sim \, {\rm e}^{-{ \frac{M_{BH}}{ T_H}}} \, \sim \,   {\rm e}^{-( M_{BH}L_P)^2} \, .
   \label{suppression0}   
\end{equation}

  Let us ask now, are all 1)-4) properties necessary for such a suppression? 
  The answer to this question is negative.  In fact  property 1),  {\it classicality}  is the defining property.  
Once we understand that a given object is classical,  the suppression of 
 any two-particle decay automatically follows, regardless of entropy and any other property.  
   In order to see this,  let us imagine that in a scattering experiment we stop short of forming a horizon,
 but still form a classical configuration of the gravitational field of the size  not that different from a Schwarzschild 
   horizon of a would-be black hole ($r_g$).   Since, there is no horizon, such a configuration 
  carries no entropy or temperature, but it is nevertheless classical. 
  
   How probable is the decay of such a classical configuration say in a two-particle quantum state? 
 Of course, intuitively we understand that the suppression must be exponential, but  in order 
 to quantify it, let us analyze classicality from field-theoretic perspective.   One of the signals of 
classicality is  that the mass $M$ of  a configuration is  much larger than its inverse size $r_*^{-1}$.  
 In other words, we deal with a coherent superposition of many soft quanta.  The number of quanta can be 
estimated to be $N \, \sim \, Mr_*$.   Now the question is how probable is the decay of a 
 coherent $N$-particle state into two quanta? This decay (up to factors of order log$N$ in the exponent) 
is suppressed as, 
  \begin{equation}
   \Gamma_{N \rightarrow 2} \, \sim \, {\rm e}^{-{N}} \, .
   \label{suppression}   
\end{equation}
 Notice,  that when applied to black holes, this counting correctly reproduces the above 
 Boltzmann (or entropy) factor, but it is {\it much more general}.  We thus learn, that neither 
 entropy nor thermality are the defining universal suppression factor, but rather classicality. 
 
 $~~~$
 
  {\it  It is classicality and not "blackholeness"  the defining universal reason for unitarization! } 
    
   $~~~$

 Being liberated from the necessity of having entropy (or even a  horizon), we  can ask  why can't 
 other classical configurations that are also result of energy sourcing  play the role  in unitarizing 
the scattering cross sections?   This is the key idea of classicalization. 
 
  In fact, we can turn the above reasoning around, and argue that classicalon configurations 
  allow us to define a notion of  an {\it entropy precursor}, in terms of the number of soft quanta 
  produced in classicalization process, which in the particular case of black holes 
  agrees with the entropy counting, but is much more general and applicable 
  to other classical configurations produced by energy self-sourcing.   This view can also shed a 
  different light at the origin of black hole entropy, by viewing it as a necessary outcome  
of classicalization, which requires production of many soft quanta, and not vice versa.

  Classicalons  share with black holes two properties that are crucial for unitarization. These 
  are,  the energy (self)sourcing and classicality.    Thus, the bare essentials of unitarization process 
 in these two different theories are surprisingly similar.   

    We hope,  that a simple calculation below illustrates  the viability and the depth of this connection. 
 For this we shall consider the  scattering process in gravity up to the point where we stop 
 short of horizon formation, and show that  formation of classical configuration is 
 closely connected to the formation of analogous configuration in a scattering process involving  
self-sourcing of  Goldstone-type scalars. 

 The idea is to repeat the calculation of the scattering process of \cite{class2},  which  was performed there for  Goldstone 
 waves,   for a  spin-2 system.   Namely, we consider scattering  of  wave-packets with a very small occupation 
number but  trans-Planckian center of mass energy and see how the system classicalizes  at larger and larger distances. 
 Not surprisingly we shall discover that the 
 role of the classicalization radius, $r_*$,  is played by the Schwarzschild radius corresponding 
 to the center of mass energy.    

This emerging parallels allow us to give an useful unified parameterization of 
  the landscape of classicalizing theories, by parameterizing  how efficiently the 
  classicalization radius $r_*(s)$  grows with energy $\sqrt{s}$.   We shall see,  that 
  gravity (spin-$2$ field)  is a most efficient classicalizer,  and  gives a linear growth, 
  $r_* \propto \sqrt{s}$. 
  In all other cases, of spin-0 classicalizer fields, the growth is slower. 
   This parameterization also allows to generalize the holographic bound on information storage 
   to all classicalizing theories. The bound is the most stringent for gravity and non-existent for 
weakly-coupled non-classicalizing theories, with spin-0 classicalizing  theories occupying intermediate states.   
    
  In order to try to achieve  a maximal clarity, we shall structure our discussion in the following way. 
 First, we shall briefly go through the essential ingredients for the Goldstone analysis, and 
 give an universal definition of the $r_*$-radius.  Then we shall repeat the similar analysis for 
 gravitons.   Finally, we shall highlight  important parallels between the two cases and emphasize  
the universal role of classicality for unitarization.

     \section{Classicalization of Goldstones} 
     
 We shall adopt the following useful definition of the classical radius $r_*$ \cite{class2}, which 
 is applicable to essentially any interacting theory, regardless of classicalization.   
 With this definition,  $r_*$ is a classical distance down to which in a high-energy scattering process the wave-packets propagate 
essentially freely, without experiencing a significant  corrections from interaction 
   terms, and beyond which the scattering can no longer be ignored. 
   Simply speaking, $r_*$ can be viewed as a distance that defines interaction range at given energy $\sqrt{s}$.
   Therefore, $r_*$ automatically  determines the cross section as a geometric  cross-section
   \begin{equation}
  \sigma \,  \sim \, r_*^2 \, . 
  \label{unisection}
  \end{equation}
  With this definition $r_*$ is a classical length that survives in the limit  $\hbar \, = \, 0$. 
   The above definition of $r_*$ is possible essentially in any interacting theory. 
    The defining property of classicalization  is the behavior of $r_*$ as a function of energy, relative to other  
quantum length-scales in the problem.    The key property for 
    classicalizing theories is, that  for $\sqrt{s} \,  \gg \, M_*$,   $r_*$ exceeds all the other quantum length 
scales in the problem. In particular, 
     $r_*(s) \, \gg \, L_*$ for  $\sqrt{s} \,  \gg \, L_*^{-1}$.  In such a system the scattering   takes place way
 before the system has any chance of probing distance $L_*$. 
   
    Following \cite{class2}, we shall now review the classicalization of scattered wave-packets  
    on a simple prototype example of  a derivatively-coupled 
   Nambu-Goldstone type scalar with the following Lagrangian,  
   \begin{equation}
 \mathcal{L}= {1\over 2}  \(\partial_{\mu}\phi\)^2 \, + \, {L_*^4\over 4}  \((\partial_{\mu}\phi)^2\)^2 \, .
 \label{nambu}
  \end{equation}
  In particular, the above Lagrangian can be viewed as a  simplest interacting  truncation of  Dirac-Born-Infeld-type theory, which is fully sufficient  for our purposes. Our results can be easily generalized 
  to higher order non-linearities (see below). 
 This theory is symmetric  under the shift by an arbitrary constant $c$,   
  \begin{equation}
  \phi \, \rightarrow \, \phi \, + \,  c \, . 
\label{goldstone}
\end{equation}   
The equation of motion is, 
 \begin{equation}
  \partial^{\mu} (\partial_{\mu}\phi \(1+ L^4_*(\partial_{\nu}\phi)^2)\) = 0 \, . 
  \label{goldequation}
  \end{equation}
In order to identify the $r_*$ radius in the above theory, let us consider a scattering 
process in which  for $r \, = \, \infty$ and $t  \, = \, - \infty$, 
$\phi$ is well-approximated by a spherical wave of very high center of mass energy 
$\sqrt{s} \, = \, A^2/a \, \gg \, M_*\equiv L_*^{-1}$ (where $a$ is a characteristic wave-length) and amplitude $A$, 
\begin{equation}
\phi_0 \, = \, {\psi(r+t)  \over r},
\label{wavepacketfree}
\end{equation}
which satisfies the free-field equation of motion,  
  \begin{equation}
  \Box\phi_0 \,   =  \, 0  \, . 
     \label{equfree}
  \end{equation}
  Since the initial wave has to describe few quantum particles, we shall assume 
$A \sim 1$ (small occupation number). 
We shall now solve the equation (\ref{goldequation})  iteratively, by representing  $\phi$-field
as superposition of a free wave-packet $\phi_0$ and a scattered wave $\phi_1$, 
\begin{equation}
\phi \, = \, \phi_0 \, + \, \phi_1\, .
\label{zeroone}
\end{equation}
We shall treat $\phi_1$ as a small perturbation and shall try to understand at what distances  
this correction to a free wave becomes 
significant. The classicalization radius will be set by a classical distance at which the approximation 
 \begin{equation}
 \phi_1 \, \ll  \, \phi_0 \,, 
 \label{scattering} 
 \end{equation}
breaks down.

 The equation for 
the leading correction to the free wave now becomes, 
 \begin{equation}
  \Box\phi_1  =  - L^4_*
     \partial^{\mu} (\partial_{\mu}\phi_0 (\partial_{\nu}\phi_0)^2) \, .
     \label{phione1}
  \end{equation}
Taking into the account properties of $\psi(t+r)$-wave,  for 
$ a \, \ll \, r$, the leading contribution to the right hand side is,  
 \begin{equation}
  \Box\phi_1  =  -  {L^4_*  \over r^5} \, ( 2 \psi^2\psi '' \, + \, 8 \psi\psi '^2) \, ,
     \label{phionea}
  \end{equation}
  where prime denotes the derivative with respect to the argument. 
For $a^{-1} \, \gg \, r^{-1}$ the solution of this equation can be approximated by, 
  \begin{equation}
  \phi_1 \, \simeq \, - f(r+t) \,  \frac{L_*^4 }{6 r^{4}} \, ,  
  \label{aprphi1}
  \end{equation}
  where,  
 \begin{equation}
 f(r+t) \, \equiv  \, \int_0^{r+t} ( 2 \psi^2(y)\psi ''(y) \, + \, 8 \psi(y) \psi '^2(y)) dy \, .
 \label{fdefinition}
 \end{equation}  
 Notice, that  since  $\psi$ is a wave-packet of amplitude $\sim 1$ and wave-length 
 $a$,   we have,   
 \begin{equation}
f \, \sim \,  \,  {1 \over a}  \psi \, . 
  \label{omegaplus}
  \end{equation}
 Thus, the breakdown of the condition (\ref{scattering}), 
 which signals that  scattering became significant, takes place  at a distance, 
 \begin{equation}
r_* \equiv  L_* ( L_*/a)^{1 \over 3}  \, .
 \label{radstar} 
 \end{equation}
 Or, translating $a$ in terms of center of mass energy, we can write,  
  \begin{equation}
r_*  \, \sim \,   L_* ( L_*\sqrt{s})^{1 \over 3}  \, .
 \label{rstars} 
 \end{equation}
  For example, for a Gaussian wave-packet,  
 \begin{equation}
\phi_0 \, =  \,  A \, {e^{-{(r+t)^2 \over  a^2}} \over r} ,
\label{wavepacket}
\end{equation}
the equation (\ref{phionea}) can be solved exactly in the limit $a \rightarrow 0$,  
 \beq
\phi_1\(r \gg a \) =-\frac{32}{9} \sqrt{\frac{\pi}{3}}  L^4_* A^3 \frac{\theta(r+t)}{a(t-r)^3 r} \, ,
\label{exact}
\eeq
which confirms (\ref{radstar}).

 Notice,  that the physical meaning of the classicalization radius can be understood also  in the following way.  
 Consider a probe source $J_{\mu_1\mu_2\, ... \mu_3}$ coupled to $\phi$. Due to the shift symmetry this coupling  
has to involve  gradients of $\phi$. For example, 
 \begin{equation}
 \left ( L_*^{2n} \partial_{\mu_1}\phi  \partial_{\mu_2}\phi ... \partial_{\mu_n}\phi  \right )  J^{\mu_1\mu_2\, ... \mu_n} \, .
\label{current}
 \end{equation}
  In  other words, the probe sees an effective background "metric" (potential), 
  \begin{equation}
  V \, = \, \left ( L_*^{2n} (\partial_{\mu_1}\phi)  (\partial_{\mu_2}\phi) ...( \partial_{\mu_n}\phi)  \right ) \, ,
  \label{potential}
  \end{equation}
  created by $\phi$, and scatters off it.  This background becomes order one at $r_*$ for $\phi_1$
 given by (\ref{exact}) and with $a \sim r_*$. In other words, a softest wave-packet 
 of a fixed energy (and thus of fixed $r_*$) would-give an order one potential  exactly at the distance  $r_*$.

 \section{Classicalization by Gravitons} 
 
   We have seen above how a  scattering of Goldstone-type scalar wave-packets 
  is classicalized due to self-sourcing.   We now wish to consider a situation in which 
  the classicalization of the same wave-packets happens through ordinary gravity. 
  
   It is well known that a collapse of a  spherical source leads to the formation of  a black hole. 
   This is well-known for the sources that are classical to start with.
   That is, when the occupation number of  the  initial wave-packet is so large that it can be treated as a classical object.    
What we wish to see instead is, that  exactly the same process for  
a small initial  occupation number  can be viewed as classicalization,
with classicalization radius $r_*$ being equal to the Schwarzschild radius ($r_g$) corresponding 
to the center of mass energy of the wave-packet.  In other words,  classicalization is  a precursor effect  in which 
initially-quantum wave-packet 
 evolves into a configuration with a classical gravitational field that  sets       
   an effective range of the interaction.

 We consider Einstein's graviton $h_{\mu\nu}$ and a massless Goldstone-type scalar $\phi$. At the level of the free  fields, 
the Lagrangian  is, 
     \begin{equation}
L_{free} \, = \, - \, {1 \over 2}  h^{\mu\nu} \mathcal{E}^{\;\;\alpha\beta}_{\mu\nu} h_{\alpha\beta} \, + \, 
{1 \over 2} \, (\partial_{\mu} \phi )^2 \, .
\label{linear1}
\end{equation}
where 
\begin{equation}
\label{einstein}
\mathcal{E}^{\; \; \alpha\beta}_{\mu\nu} h_{\alpha\beta}\, = \, \Box h_{\mu\nu} \, - 
\,\eta_{\mu\nu}  \Box h  \, - \, \partial_{\mu}\partial^{\alpha} h_{\alpha\nu} \, -\, 
\partial_{\nu} \partial^{\alpha} h_{\alpha\mu} \, + \, 
\eta_{\mu\nu} \partial^{\alpha}\partial^{\beta}h_{\alpha\beta}\, + \, \partial_{\mu}\partial_{\nu} h
\end{equation}
is a linearized Einstein tensor. 
 Eq (\ref{linear1})   represents an unique linear  ghost-free action for a massless spin-2 field. 
     This theory propagates  2 degrees of freedom, and is invariant under the following shift,           
    \begin{equation}
\delta h_{\mu\nu} \, = \, 
\, \partial_{\mu} \xi _{\nu} \, + \, \partial_{\nu} \xi _{\mu}  \label{hdecomp} \, ,
\end{equation}
where $\xi_{\mu}$ is an arbitrary vector  (below we shall use harmonic gauge $\partial^{\mu} h_{\mu\nu} \, = \, {1 \over 2} \partial_{\nu} h$).
 In order to see how graviton classicalizes in the scattering process, we shall introduce 
 the  interaction terms.  For our purposes we shall limit ourselves by cubic order interactions in fields.  To this order the  coupling can be accounted by coupling graviton to the energy momentum tensor of the  system. That is,  we supplement (\ref{linear1}) by,  
   \begin{equation}
L_{interaction} \, = \, L_P \, h_{\mu\nu}  \,  T^{\mu\nu}(\phi)   \,  +  \,  L_P \, \mathcal{O} (h^3)_{Einstein} 
 \label{coupling}
\end{equation}
where,  $L_P \equiv M_P$ is the Planck length, and 
 \begin{equation}
 T_{\mu\nu}(\phi) \, = \, \partial_{\mu}\phi \partial_{\nu}\phi \, - \, {1 \over 2}\, 
 \eta_{\mu\nu}
 (\partial_{\alpha}\phi \partial^{\alpha}\phi\,) \, 
 \label{scalarenergy}
 \end{equation}
and  $\mathcal{O} (h^3)_{Einstein}$, stands for a cubic part of the Einstein's action, which
can be thought of as a self-sourcing of graviton by its own energy-momentum tensor. 
 For simplicity, we  shall not write this term explicitly in the action, but only its contribution into the equation of motion.  To this order (and using harmonic gauge)  the equation of motion for the graviton has the following form,   
  \begin{equation}
  \square  \, h_{\mu\nu} \, = \, - \, L_P \, (T_{\mu\nu} \, - \, {1 \over 2} 
  \eta_{\mu\nu} \, T_{\alpha}^{\alpha} \, ),
   \label{gravitoneq}
  \end{equation}
 where 
    \begin{equation}
 T_{\mu\nu}\, \equiv  T_{\mu\nu}(\phi) \, + \, T_{\mu\nu} (h) \, 
\label{tmunu}
\end{equation}
 and  
  \begin{eqnarray}
   T_{\mu\nu} (h) =&&
 - {1\over 2} \, h^{\alpha\beta} \left ( 
 \partial_{\mu}\partial_{\nu}  h_{\alpha\beta} \, + \, \partial_{\alpha}\partial_{\beta}  h_{\mu\nu}
 \, -  \partial_{\alpha}(\partial_{\nu}  h_{\mu\beta}  \, + \, \partial_{\mu}  h_{\nu\beta})
 \right ) \, - \nonumber \\
&&  - {1 \over 2}  \partial_{\alpha} h_{\beta\nu} \partial^{\alpha} h^{\beta}_{\mu} 
   +    {1 \over 2}  \partial_{\alpha} h_{\beta\nu} \partial^{\beta} h^{\alpha}_{\mu}  \,
   -\, {1 \over 4} \, \partial_{\mu} \, h_{\alpha\beta} \, \partial_{\nu} \, h^{\alpha\beta} 
 \nonumber \\
 &&    
   - \, {1 \over 4} \, \eta_{\mu\nu} \, \left ( 
   \partial_{\alpha} \, h_{\beta\gamma} \, \partial^{\beta} \, h^{\alpha\gamma}  \, - 
   \, {3 \over 2} \,  \partial_{\alpha} \, h_{\beta\gamma} \, \partial^{\alpha} \, h^{\beta\gamma} )
   \right )  
   \nonumber \\
   &&
        \, - \, {1\over 4} h_{\mu\nu} \, \square \, h \, + \,  
  {1\over 2} \,  \eta_{\mu\nu} \, h_{\alpha\beta} \, \square \,  h^{\alpha\beta}\, .
  \label{gravitonsource}
 \end{eqnarray} 
 
  Since, we have already discussed the classicalization due to self-sourcing of $\phi$, we shall not 
include the self-interaction terms for simplicity, but only terms that source graviton.   
  
  We shall now proceed as follows. We assume that our system starts 
  at $t \, =\,  -\, \infty$ and $r \,  = \, \infty$  
  in an "in"-state with no gravitons and a collapsing free wave-packet of 
  $\phi \, = \phi_0$ with a trans-Planckian  center of mass energy and a small occupation number. 
  
   Perturbatively, such a $\phi$-scattering due to graviton exchange would   violate unitarity 
for $\sqrt{s} \, \gg \, M_P$.    Instead we shall see that system classicalizes via production of  
a large gravitational field.  
  
  We shall observe how the initial quantum wave will evolve into a classical configuration of the graviton field, 
and how the classicalization distance will depend on energy.  In order to achieve this goal,  we shall look for the 
solution for graviton in form of an expansion, 
  \begin{equation}
h_{\mu\nu} \, = \,  h^{(0)}_{\mu\nu} \, + \,  h^{(1)}_{\mu\nu} \, ,  
 \label{hexpansion}
  \end{equation}
 where  $h^{(0)}_{\mu\nu}$  is the solution of the linear equation,  that takes into the account sourcing of graviton by 
 $T_{\mu\nu}(\phi)$ (but ignores self-sourcing!),  whereas, 
 $h^{(1)}_{\mu\nu}$  is the next order correction due to self-sourcing by  
 $T_{\mu\nu} (h)$.   We than try to find out at what radius,  
 \begin{equation}
 h^{(0)}_{\mu\nu} \sim  \,  M_P
 \label{cond1}
 \end{equation}
  and 
  \begin{equation}
  h^{(1)}_{\mu\nu} \,  \sim 
  h^{(0)}_{\mu\nu}  \, .
   \label{cond2}
 \end{equation}
This will define the classicalization radius, $r_*$. 
   We shall discover that this distance is set by the classical scale 
  equal to a Schwarzschild radius corresponding to a center of mass energy. 
   This is a clear signal that system classicalizes by formation of classical objects 
   whose size is governed by the Schwarzschild radius.  
   Let us now analyze the scattering process.  We choose,   $\phi_0$ to be a free spherical wave-packet, 
 \begin{equation}
\phi_0 \, = \, {\psi(r+t)  \over r} \, ,
\label{wavepacket0}
\end{equation}
of amplitude  $A$ and energy $\sqrt{s} \, \equiv \, M \sim A^2/a$, where $1/a$ is a characteristic momentum  
(or inverse localization width) entering in  the wave-packet. 
For evaluating $h^{(0)}_{\mu\nu}$ we need to solve equation  (\ref{gravitoneq}) with 
the source  (\ref{scalarenergy}), which takes the following form, 
  \begin{equation}
  \square  \, h_{\mu\nu}^{(0)} \, = \, - \, L_P \, ( \partial_{\mu}\phi_0 \partial_{\nu}\phi_0 ) \, .
   \label{gravitoneq1}
  \end{equation}
  Let us evaluate this equation on a spherical  wave-packet. Picking up (a most interesting) 
 $h_{00}$  component of the graviton that  in classicalized  limit should reproduce Newtonian potential of a classical 
source,  we get 
   \begin{equation}
  \square  \, h_{00}^{(0)} \, = \, - \, L_P \, {\psi'^2 \over r^2} \, ,
   \label{gravitoneq2}
  \end{equation}
   where prime denotes derivative with respect to the argument (in this case $(t+r)$). 
  Is is obvious that for any localized wave-packet  with total center of mass energy  $M$, outside the source 
  the Newtonian gravitational potential must behave as $ h_{00}^{(0)} \sim  L_PM/r$, 
  which makes it clear that the classicalization radius is given by the Schwarzschild radius 
  associated with the  center of mass energy $M$.  
  
    To make the analogy with the Goldstone case deeper,  we wish to represent the metric in the form that is most 
convenient for confronting 
   it with the case of a classicalized scalar field (\ref{exact}).  For this,  let us take  
 \begin{equation}
\phi_0 \, = \, {\psi(r+t)  \over r} \, = \,  A \, {e^{-{(r+t)^2 \over   2a^2}} \over \pi^{1/4} r} ,
\label{wavepacket1}
\end{equation}
which represents a wave-packet  of amplitude  $A$ and energy $M \sim A^2/a$.    
The equation (\ref{gravitoneq2}) now becomes (irrelevant factors are absorbed in $L_P$), 
   \begin{equation}
  \square  \, h_{00}^{(0)} \, = \, - \, 2\,  L_P \, M {1 \over r^2} 
  \left ( 1 +  {1\over 2} a^2 \partial_t^2 \right ) \, {e^{-{(r+t)^2 \over   a^2}} \over a \sqrt{\pi}} 
   \label{gravitoneq3}
  \end{equation}
  For clarity of the solution, we shall take a limit  $a\to 0$,   $A^2/a =$fixed, and  use the relation 
\beq
\lim_{a\to 0} {1 \over a \sqrt{\pi}} e^{-\frac{(r+t)^2}{2a^2}} \, = \, \delta(r+t) \,  .
\eeq
 The equation then becomes, 
    \begin{equation}
  \square  \, h_{00}^{(0)} \, = \, - \, 2 \, L_P \, {M \over r^2}  \delta(r+t) \, ,
   \label{gravitoneq4}
  \end{equation}
which gives,
     \begin{equation}
   h_{00}^{(0)} \, =  \, { L_P M \over r}   \, \theta(r+t) {\rm ln} (r-t) \, .
   \label{gravitonforml}
  \end{equation}
   Comparison of  this expression with (\ref{exact}) suggests a clear analogy between the two cases.  In particular,  
it is obvious that in gravity $L_P$ plays  the role of $L_*$,   
 whereas the role of $r_*$ is played by the Schwarzschild radius  associated with energy 
 $\sqrt{s} \, \equiv \, M$.  
Indeed,   classicalization radius is set by the condition   (\ref{cond1}), which is reached at 
  $r_* \sim  L_P^2M$,  and which  is nothing but the Schwarzschild radius  of  a black hole of mass 
  $M$.   
  
   Note, that  appearance of a time-dependent log-factor in (\ref{gravitonforml}), is an artifact of the 
  harmonic gauge we are working in.  In this gauge,  appearance of the log-factors in the 
  gravitational tails at  $1/r^2$-order in source expansion is a well known feature in gravitational wave
  physics, and can be removed by appropriate gauge shift, which effectively amounts to a correction of  
light-cone outside the source  \cite{waves},\cite{gravitywaves}.   However, for our purposes, of making the parallel between the 
graviton and Goldstone classicalization  more transparent, we
  prefer to stay in the above gauge.  
  
   Now   evaluating  (\ref{gravitonsource}) on (\ref{gravitonforml}) and 
  inserting it  eq (\ref{gravitoneq}) we can get the standard corrections to the metric at second order in $L_P^2$. For example, 
  \begin{equation}
   {h_{00}^{(1)} \over M_P}  \, \sim  \,  {1 \over 2} \, {r_*^2 \over r^2}  \, , 
     \label{bilineargraviton}
    \end{equation}
which clearly indicates that the second condition of classicalization  (\ref{cond2}) is also 
met at $r\sim r_*$.  We thus see, that  the classicalization radius for spin-2 coincides  
with the Schwarzschild radius corresponding to the center of mass energy. For the interested reader, there is 
a full non-linear treatment for the collapse of a spherical light shell confirming our findings for the formation
of the classicalon (black hole in this case) \cite{lightshell}, for us however most important is the 
deconstruction of the collapse from classicalization point of view, as we did above.

\section{Classicalization of Graviton Scattering}

  We shall now study the classicalization of graviton scattering. For this we shall repeat the same exercise as 
in the previous case, but replace an incoming scalar wave-packet by 
  a graviton one.   We shall thus again solve the equation (\ref{gravitoneq}) iteratively, by 
  setting the scalar field to zero, and taking $h_{\mu\nu}^{(0)}$ as a free incoming gravitational 
  wave-packet.    All the machinery required for such analysis has already been prepared 
  in series of excellent papers \cite{waves},\cite{gravitywaves},\cite{T2} studying both the multipole and the post-Newtonian 
  expansions of gravitational waves.   We shall limit ourselves by adapting  the key essentials of this analysis
 for our purpose of understanding classicalization of scattered gravitons.

     For standard convenience  we rewrite the equation (\ref{gravitoneq}), in notation  
    of trace-reversed graviton,  
     \begin{equation}
\bh_{\alpha\beta}=h_{\alpha\beta}-\frac{1}{2}\eta_{\alpha\beta}h^\kappa_\kappa\, , 
\end{equation}
which in harmonic gauge  $\partial_\mu \bh^{\mu\nu}=0$ takes the form, 
\begin{equation}
\Box \bh^{\mu\nu}=- \frac{1}{M_P} \Lambda^{\mu\nu} \label{eqh}
\end{equation}
where $\Lambda^{\mu\nu}$ is given by
\begin{eqnarray}
  \Lambda_{\mu\nu} (\bh)&\!\!=\!\!& 
-\bh^{\kappa\lambda}\p_\mu\p_\nu \bh_{\kappa\lambda}+\p_\kappa \bh^{\mu\lambda}\p_\lambda \bh^{\nu\kappa}+\frac{1}{2}
\eta^{\mu\nu} \p_\lambda h^{\sigma}_{\kappa}\p_\sigma \bh^{\kappa\lambda}\nonumber \\
&&-\p_\lambda \bh^{\nu\tau}\p^\mu \bh^\lambda_\tau-\p_\lambda \bh^{\mu\tau}\p^\nu \bh^\lambda_\tau+\p_\lambda 
\bh^{\mu\tau}\p^\lambda h^\nu_\tau \label{Lambda}\\
&&+\frac{1}{8}\left(2 \eta^{\mu\lambda}\eta^{\nu\kappa}-\eta^{\mu\nu}\eta^{\kappa\lambda}\right)\left(2\p_\kappa \bh^{\tau\sigma}
\p_\lambda \bh_{\tau\sigma}-\p_\kappa \bh^{\tau}_\tau\p_\lambda \bh^\sigma_\sigma\right)\, . \nonumber
 \end{eqnarray}    
  We shall solve this equation iteratively by representing a graviton as a superposition
  of free and scattered waves,     
  \begin{equation}
h_{\mu\nu} \, = \,  h^{(0)}_{\mu\nu} \, + \,  h^{(1)}_{\mu\nu} \, ,  
 \label{bhexpansion}
  \end{equation}
which to the leading order satisfy the following equations, 
\begin{eqnarray}
\Box \bh_{\mu\nu}^{(0)}\, = \, 0\, , ~~~~ \partial_\mu \bh_{\mu\nu}^{(0)} \,  = \, 0
\label{freeh}
\end{eqnarray}
and 
\begin{equation}
\Box \bh_{\mu\nu}^{(1)} \, = \, - \,  \frac{1}{M_P} \Lambda^{\mu\nu}(h^{(0)}) 
\label{eqhpert}
\end{equation}
The solution to (\ref{freeh}) is usually given in terms of infinite multipolar series using symmetric trace-free harmonics \cite{T2}
and takes the form of multipolar waves 
\begin{eqnarray}
\bh^{(0)\mu\nu}=\sum_{\ell=0}^\infty \partial_L \left(\frac{K_L^{\mu\nu}(u)}{r}\right)
\end{eqnarray}
Here $L=i_1i_2\ldots i_\ell$ is a compact notation for multipolar index of $\ell$ spatial indices. Thus, $K_L^{\mu\nu}=
K_{i_1i_2\ldots i_\ell}^{\mu\nu}$ and $\p_L=\p_{i_1}\p_{i_2}\ldots \p_{i_\ell}$. In particular $K_{i_1i_2\ldots i_\ell}^{\mu\nu}$
 in our case is a function of  $u\equiv t+r$ and it is symmetric and trace-free in its lower indices.   

  For simplicity we shall choose incoming wave to be a quadrupole,   
\begin{eqnarray}
&&\bh_{00}^{(0)}=-2\partial_{i}\p_{l}\left(\frac{M_{il}(u)}{M_P\, r}\right)\, ,  \nonumber \\
&&\bh_{0i}^{(0)}=2\partial_{j}\left(\frac{M_{j \, i}^{(1)}(u)}{M_P\, r}\right)\, ,  \\
&&\bh_{ij}^{(0)}=-2\frac{M_{ij}^{(2)}(u)}{M_P\, r}\, ,  \nonumber \\
\label{quadrupole}
\end{eqnarray}
where $M_{kl}^{(n)}=d^nM_{kl}/dt^n$. 

 The fact, that very similar classicalization mechanism is at  work as in the case of scalar scattering, can already be 
anticipated from the observation, that when evaluated on (\ref{quadrupole}) 
 to the leading order of $1/r^2$ contribution, the right hand side on the equation 
 (\ref{eqhpert}) takes the form of the stress energy tensor of a massless field, 
  \begin{equation}
\Box \bh_{\mu\nu}^{(1)} \, = \, - \,  \frac{1}{M_P} {1\over r^2}  k_{\mu} k_{\nu} \Pi(t+r)
\label{eqhpert1}
\end{equation}
where, $k_{\mu}$ is null vector, and quantity $\Pi$ accounts for the multipole structure of the product of the two quadrupoles, 
and can be re-expanded in the standard way.
 
   In order not to stress the reader with a tedious but straightforward calculation 
 of general case (which can be found in standard gravitational wave  analysis \cite{waves},
\cite{gravitywaves},\cite{T2}), we limit ourselves by illustrating the point for
 a monopole component of $\Pi(t+r)$, in which case the problem essentially reduces to a scalar 
 incoming wave studied in the previous section.  So the gravitational $h_{\mu\nu}^{(1)} $ field produced  by 
a monopole component  of the self-source is  now of the same nature as the 
 field $h_{\mu\nu}^{(0)}$ produced by sourcing by a  scalar wave in the equation (\ref{gravitonforml}). 
   For a sharply localized incoming free wave-packet,  the $h_{00}^{(1)}$ 
  produced by a monopole component of a self-source can be approximated by  
  \begin{equation}
   h_{00}^{(1)} \, \sim  \,  \, { L_P \sqrt{s} \over r}   \, \theta(t+r) {\rm ln} (r-t) \, , 
   \label{gravitonforml1}
  \end{equation}
where $\sqrt{s}$ is (a spherically averaged) center of mass energy of an incoming wave. 
 Just as in the scalar case, this fixes the classicalization radius at  the Schwarzschild radius corresponding 
to this center of mass energy $r_* \sim r_g = \sqrt{s}L_P^2$.

\section{Parallels and Differences} 

\subsection{Classicalons in Gravity versus Goldstone} 

We have proven the emergence of the classical radius $r_*$ in the scattering process,  
both for gravitons as well as Goldstones.  In both cases this classical radius marks  an effective range of 
interaction, that is, a distance where a free-wave approximation breaks down.  In both cases a correction to a free
wave is leaving behind a power-law tale. All the above signals,  that we deal with a configuration that became classical 
at $r_*$.  Thus, whatever the subsequent evolution of the system is,  an $s$-wave scattering 
of highly energetic quanta leads to a formation of a classical field configuration of size $r_*$. 
In neither cases this classicalon configuration yet corresponds to any {\it static}  spherically-symmetric  solution.  
 
   What is a post-classicalization  evolution of  fields? 
  
  In case of gravity the answer is simple.   Classicalization must be a precursor of the formation of 
  a Schwarzschild black hole.  This is obvious from the spherical symmetry of the problem and 
  from Birkhoff's theorem.  Of course, the details of horizon crossing cannot be captured by our analysis, but 
  there is no need for it.  Our goal was to witness classicalization. 
  
   What is an analogous static solution (if any) in the scalar case to which the classicalon configuration may settle? 
As found in \cite{class1}, theory does admit such static singular solutions, which in some respect are analogous to black holes. 
Whether  a time-dependent classicalon configuration 
   will settle to this, is unclear. But, the key point is,  that from the unitarization point of view there 
   is no need for a settlement to any stable or long-lived solution, since classical configuration 
   has already being formed and this is enough.   Once formed, the classical configuration 
   cannot decay in a two-particle state without paying price of exponentially-suppressed probability. 
  
   This simple but important  fact,  naturally allows us to resolve  the question of the role of  black hole-type entropy 
for unitarization of the scattering amplitude.  Since black hole entropy  is a property  of very special classical 
configurations, whereas any intermediate classicalon  that is 
  formed by self-sourcing   suffices  to unitarize the scattering  process, it follows that black hole entropy  is just an 
accompanying property for gravitational systems  (which of course automatically agrees with the counting of 
exponential suppression in two-particle decays)  but not a defining  property for unitarization.

   \subsection{The role of entropy in unitarization by classicalization}
   
   The closed parallel presented here between gravity and Goldstone examples, clarifies a potential 
   misconception about the role of black hole entropy in unitarization process.  
    We wish to discuss this point in more details. 
    The process of unitarization by classicalization  consists of two ingredients. 
    
    $~~~$
    
   {\bf (a)} First is to show, that an initial two-particle state inevitably leads to the development of classical
    configuration.  We have demonstrated this for an $s$-wave scattering. 
    
    $~~~$
    
   {\bf (b)}  The second ingredient is the fate of this classical configuration:  An exponential suppression of  its  
decay-probability  into a two-particle state. 
     
   $~~~$  
     
      Let us investigate the role of black-hole-type entropy for any of these stages.  Obviously, entropy plays 
      no role at the first (classicalization) stage neither in Goldstone nor in gravity examples. 
     We saw, that  classicalization took place before one could talk about any entropy or even a well-defined horizon. 
      
       What role does entropy play in the second aspect?  In gravity example, entropy is 
       a byproduct of classicalization, since we know that a spherical wave packet after collapsing 
       will form a Schwarzschild black hole with an associated Bekenstein-Hawking  entropy. 
       So in gravity, at least in the $s$-wave scattering,  the outcome of classicalization is a black hole formation, 
and this is accompanied by formation of entropy.  However,  main reason for unitarization 
       is classicality,  not the entropy of the final classical configuration.  Any classical configuration, with or without entropy,  can only decay into  a 
two-particle state with exponentially-suppressed probability. 
   Since entropy is a inevitable accompanying property of classicalons  in gravity case,  entropy counting does agree 
with this suppression. But classicality is the key reason 
 for the suppression regardless of entropy. 
    For example, a binary system of neutron and anti-neutron stars carries no black hole type 
    entropy, but probability for it to decay into two photons is obviously suppressed. 
    
     This is a defining property of all classical configurations. They represent many particles in 
     coherent state and due to this their decay into any two-particle states is exponentially 
     suppressed, by a factor $e^{-Nc}$, where $N$ is number of soft quanta composing the configuration in question, $c$ 
is a factor  of order one  up to $log N$ corrections. An exact computation of this suppression factor  is beyond the 
scope of our paper, but  it  is very instructive 
     to estimate it from the following reasoning.  
     
       The classicalon configuration of size  $r_*$ and mass $\sqrt{s}$ is mostly composed out of soft quanta of 
wave-length $\sim r_*$. Therefore their  number  in a classicalon configuration 
       can be estimated as  $N \sim \sqrt{s}r_*$,  and thus the decay into two particles is suppressed 
  by a factor, 
  \begin{equation}
        \Gamma_{classicalon \rightarrow few} \sim {\rm e}^{-\sqrt{s}r_*} \, .
        \label{classrate}
        \end{equation}
    Notice that for the gravity case  ( $r_* \,  =\, r_g \, =\,  \sqrt{s}L_P^2$) 
   the above expression correctly reproduces entropy suppression of a black hole decay into 
   few particles,  although we never referred to any entropy!

 \subsection{If not $N \rightarrow  2$,  then why $2\rightarrow N$?}
 
  We now wish to address a question that may bother some readers. 
  
  We have just argued that two-particle decay of any classical configuration,  which can always be understood as a 
superposition of many quanta,  is exponentially suppressed. 
   In this  light how can we understand an unsuppressed formation of the same classical  
 state out  of initial few energetic quanta? 
 
  One may think that in the case of a black hole the large entropy factor somehow 
 plays the role and overcompensates. This way of counting is misleading.  First,  as we know, the gravitational 
 field classicalizes before horizon is even formed and one can prescribe standard 
 Bekenstein-Hawking  entropy. 
 The two-body decay of such a classical field would be suppressed even if it never 
 evolves into a black hole.   Moreover, the origin of the black hole's entropy 
 is precisely the result of the fact that many initial states evolve into the same final one, and not other way round.   
 
 The fact that the same black hole can be formed by many different initial states, 
gives no enhancement for black hole formation from each initial pure state.   
 
   So the answer  to the above question has nothing to do with the particularities of the  black hole entropy,  but rather again is universal property of classicallization.   
 To answer this question let us notice, that  when we form a classical configuration 
 out of an initial wave-packet describing free particles, we are dealing with a transition 
 that takes infinite time, because  initial particles can be considered as free only if 
 $r\, \gg \, r_*$.   
 
   So the $N$-particle classical configuration is built up gradually,  during a very long time, 
   and at distances at which the quantum re-scattering is negligible. 
   Of course, there is full time reversal symmetry, and $N$-particle state could 
   gradually bounce back into the original $2$-particle state, however , the time needed 
   for this is much longer than the time of re-scattering among the $N$-particles, which takes away 
   the needed coherence. As a result, smooth bounce back almost never happens.  
   
    Also, since classicalization (both for gravity and other cases) forces the system to produce 
    $N$ soft quanta, the enhancement can only be encoded in the number of states in which 
    these particles can be produced, and must be universally applicable for gravity and for other cases. 
   
     Let us analyze this discussion in an universal field theoretic language, equally applicable both 
   for black holes and as well as for  other classicalons.   Consider an above-considered  scattering process in a 
theory (not necessarily a classicalizing one)  in which the interaction range  $r_*(s)$ is some function of center of 
 mass energy $\sqrt{s}$.   
       This means that highly energetic wave-packet starts to scatter at  distance $r_*$. 
    Typical quanta produced in this scattering have wave-length $r_*(s)$. Their available number can be estimated as  
$N = \sqrt{s} r_*(s)$, for $r_*(s) \, \geq \,  1/\sqrt{s}$ 
  and $N \sim 1$ in the opposite case  $r_* \, \leq \, 1/\sqrt{s}$. 
    .  Let us ask, what is the suppression factor for producing a composite object of the available mass $\sqrt{s}$ 
composed out of quanta 
    of wave-length $\bar{r}$. The number of quanta in such an object will of course be 
    $\bar{N} \, \equiv  \, \sqrt{s} \bar{r}$.   Up to log-corrections in the exponent,  the exponential suppression
 factor for such a process will be 
  \begin{equation}
   {\rm e}^{-{ \bar{r} \over r_*(s)}} \, \equiv \,     {\rm e}^{-{\bar{N} \over N}} \, .
    \label{exprice}
    \end{equation}
   Notice that this universal language correctly reproduces the suppression price for soliton production in high energy
 scattering in weakly coupled (non-classicalizing)  theories, because in such theories  at high  energies  
$r_*(s)  \, \ll \, 1/\sqrt{s}$, whereas, because  size of the soliton $\bar{r}$ is much larger than its inverse mass  
$\bar{r} \, \gg \, 1/\sqrt{s}$, hence an exponentially 
   suppressed price.  For example, for t'Hooft-Polyakov monopoles,  in a weakly coupled  theory with gauge-coupling $g$, 
 the monopole size  and mass in terms of 
 gauge boson mass  are given  by 
$1/m_{W}$ and $m_{W}/g^2$ respectively.   The minimal center of mass energy for pair production is 
$\sqrt{s} \, \sim \,  m_{W}/g^2$.   
   Thus, we have $\bar{r} \sim (1/(\sqrt{s}  g^2))$,  whereas $r_*(s) \, \sim \, g^2/\sqrt{s}$.
  Hence for a  production of monopole anti-monopole pair in scattering of two particle we obtain the following suppression, 
      \begin{equation}
  \Gamma_{2 \, \rightarrow \, mon. + anti-mon.} \, \sim \,  {\rm e}^{-{ 1 \over g^2 }} \, , 
     \label{monopole}
    \end{equation}
 which very well reproduces the expected suppression for production of solitonic objects. 
   However, when applied to classicalons the same method  shows that suppression disappears. 
   This is because, for classicalizing theories $r_*(s) \, \gg \, 1/\sqrt{s}$ and 
   $\bar{r} \, = \, r_*(s)$. So the exponent becomes one!  This  explains why production 
   of black holes is unsuppressed, without ever referring to any entropy.
   
    The physics of suppression  is entirely clear  in terms of soft quanta.   The rule is simple. 
   Suppression in a production of a classical object of size $\bar{r}$ appears whenever the number of soft quanta 
required for making up this object ($\bar{N}$) is  larger than the number 
 of soft quanta $N$ that the system can produce at the distance $r_*(s)$.  
 For weakly coupled solitons $\bar{N} \, \gg \, N$, and the suppression is given by (\ref{monopole}).

 For black holes and other classicalons, these two numbers are similar, 
$\bar{N} \, \simeq \,  N$  
 and the suppression disappears.  In particular case for black holes, $r_*(s)$  is the Schwarzschild radius and  
$N \sim r_g\sqrt{s}$, which is the same as entropy, but the physics is much more general, and what matters is 
simply the number of soft quanta produced by scattering.

  Having understood in this language why  classicalon formation avoids any price of exponential 
  suppression  that one would naively expect, let  us now understand the suppression of the 
  two particle decay in the same language.  This suppression  is simply coming 
  from the fact that  $N$ soft quanta that compose the classicalon have to annihilate into 
  $2$ very energetic ones.  This obviously costs (\ref{classrate}), which for the black hole 
  automatically reproduces the entropy or Boltzmann suppression, but there is no need 
  to refer to any of these notions.  As we see the answer can be understood in very general 
  terms of need to annihilate many soft  quanta into the two energetic ones.  
  
  There no violation of time reversal symmetry or any other exotica in physics of classicalons. 
 Rather the physics is simple. Production of classicalon  in high energy collision is easy     
 because in classicalizing it is easy to produce many soft quanta  out of enough energy. 
 However, once soft quanta have been produces their annihilation into few energetic ones is 
 always  costly.

\section{Classicalization landscape: Gravity as a most  efficient classicalizer}
\subsection{Graviton as a most efficient classicalizer} 

 In our treatment of the scalar example,  the effect of classicalization on the scattering 
 amplitude can also be represented  in the following terms.   We have started with a theory  in which 
 at low energies ($\sqrt{s} \ll  L_*^{-1}$) an $s$-channel $2\rightarrow 2$ scattering amplitude behaves 
 as, 
 \begin{equation}\label{one}
A(s)|_{s \ll L_*^{-1}} \sim (L_*\sqrt{s})^4 \, .
\end{equation}
Naive extrapolation of the above  perturbative amplitude at high energies would imply 
violation of unitarity for $\sqrt{s} \, \gg \, L_*^{-1}$. Instead, the amplitude behaves  as, 
\begin{equation}\label{two}
A(s)|_{s\gg L_*^{-1}} \sim \left (\frac{L_*}{r_*(s)} \right )^4 \, , 
\end{equation} 
 with $r_*(s)$ being given by (\ref{rstars}).   This is the key idea of unitarization 
 by classicalization.   Since, the high energy scattering takes place at  the classical radius 
 $r_*$, two-to-two scatterings are dominates by momentum-transfer  of order $ r_*(s)^{-1}$.
 In a generic  classicalizing theory the $\sqrt{s}$-dependence of $r_*$ can be parameterized   as, 
 \begin{equation}
 r_*(s) = L_*(\sqrt{s}L_*)^{\alpha} \, , 
 \label{rstaralpha}  
\end{equation}
where for classicalization  it is necessary that $\alpha$ is a positive number. 
 Its precise value  depends on the self-sourcing interaction.   As we have seen,  for a graviton
 $\alpha = 1$.   
 
   Let us now show, that  in Poincare-invariant  theories  with a scalar classicalizer field,  $\alpha$ has to be below one.  
Following \cite{class1},  consider a (self)sourcing by an operator that contains $2k$ power of derivatives  and $n$ power of a scalar field $\phi$,  
which schematically can be written as 
   \begin{equation}
   G_{\phi} \partial^{2k} \phi^n \, ,
   \label{operator}
   \end{equation}
 where $G_{\phi}$ is a coupling constant of dimensionality $G_{\phi} = [mass]^{{2-n \over 2}}
[length]^{{n + 4k -6 \over 2}}$.    This coupling defines a quantum length-scale 
$L_* \, = \, G_{\phi}^{{1 \over n+2k -4}}  \hbar^{{n-2 \over 2(n+2k-4)}}$ that marks the 
breakdown of perturbative unitarity.   
 
    The classicalization radius caused by such an operator,  can be estimated by noticing that  the only classical 
length that  can be constructed out of  the scale $L_*$ and a 
 de Broglie wave-length of the wave-packet  $a \, = \, {\hbar \over \sqrt{s}}$ is 
 \begin{equation}  
   r_* \, = \, L_* \left ({L_* \over a} \right)^{{n-2 \over n+4k -6}} \, .
   \label{rclass}
   \end{equation}  
 Of course, the above dimensional analysis does not prove that the above operator always leads to classicalization,  
but whenever it does, the classical radius scale is uniquely  defined
 and since $k \geq 2$  ($k=2$ corresponds to a free theory by field redefinition), we have 
$\alpha =   {n-2 \over n+4k -6}  \, < \, 1$.
 
 For the purposes of our present discussion, more reliable proof that 
 $\alpha < 1$,  is to show that  the length scale at which the wave-packet first gets perturbed by 
 self-interaction, which therefore must always be equal or larger than $r_*$,  must scale 
 as fractional power of  $1/a$.   
 
    Let us estimate such   
  radius  caused by such an operator  in the above-considered scattering process. 
   The self-sourcing equation (analog of (\ref{phione1}))  can be schematically written as,   
   \begin{equation}
  \Box\phi_1 \,  = \,   L_*^{2k +n-4}
     \partial^{2k}  \phi_0^{n-1} \, , 
     \label{phionegeneral}
  \end{equation}
 where $\phi_0$ is an initial  free wave-packet (given by (\ref{wavepacketfree})) of characteristic wave-length $a$ 
and energy $\sqrt{s} \sim 1/a$.    The ordering of derivatives and fields is model-dependent and is  not  shown explicitly.   
Because $\phi$ is a scalar,  all the Lorentz-indexes of derivatives on the r.h.s. of the equation
 (\ref{phionegeneral}) are contracted  among each other.   Due to this fact,  for 
 $a \, \ll \, r$  each pair of  contacted derivatives results into a factor $1/(ar)$ as opposed to 
 $1/a^2$ that one would naively anticipate.  So in total the effect of derivatives is crudely reduced 
 to  an appearance of a factor of order   $1/(ra)^n$.  Thus generically,  the equation will have a form, 
    \begin{equation}
  \Box\phi_1  \,  \sim \,   L_*^{2k +n-4} \, 
     {1 \over a^k}  \, {f(r+t) \over  r^{n + k-1}} \, , 
     \label{phionegeneral1}
  \end{equation}
 where, $f(r+t)$ is a wave-packet  of wave-length $a$ and of order-one amplitude.   
  In order to solve for $\phi_1$ we have to invert the box, which effectively removes one power of $1/(ar)$, 
so that finally we have,  
    \begin{equation}
  \phi_1  \,  \sim \,   L_*^{2k +n-4} \, 
     {1 \over a^{k-1}}  \, {\tilde{f}(r+t) \over  r^{n + k-2}} \, , 
     \label{phionegeneral2}
  \end{equation}
   where, 
    \begin{equation}
 \tilde{f}(r+t) \, \equiv  \, {a}  \int_0^{r+t} f(y) dy \, .
 \label{ftilde}
 \end{equation}  
 Since, both $\psi$ and $\tilde{f}$ have order-one amplitudes, the $\hat{r}$ radius, which marks  the  breakdown of the 
free-wave approximation (\ref{scattering}) ,  is given by 
 $ \hat{r} \, \sim \, L_* (L_*/a)^{(k-1)/(n+k-3)}$. Or, translated in terms of center of mass energy of the 
 initial wave-packet ,  we can write, 
 \begin{equation}
 \hat{r} \, \sim \, L_* (L_* \sqrt{s})^{(k-1)/(n+k-3)} \, . 
 \label{rstargood}
 \end{equation}
 The key point is,  that by default  for any system $\hat{r}$ represents an upper bound on $r_*$, since system cannot 
classicalize before experiencing  interaction. 
 Notice, also that for $k=n/2$,  $\hat{r} = r_*$. 
 
      It may be instructive to see this more transparently by taking a limit of a sharply localized wave-packet,   
$\phi_0 \, = \,  A\,  e^{-\frac{(r+t)^2}{a^2}} $.    Taking now a limit $a \rightarrow 0$ 
      and $\sqrt{s} \sim A^2/a = $fixed,  the function  $f (r+t)$  in 
   (\ref{phionegeneral1}) becomes $\delta(r+t)$,    
     \begin{equation}
 \Box \phi_1  \,  =  \,   L_*^{2k +n-4} \, 
    {A^{n-1} \over a^{k-1}}  \, {\delta(r+t) \over  r^{n + k-1}} \, , 
     \label{phionegeneral4}
  \end{equation}
  (here and below irrelevant numerical factors will be dropped).     
  Which can be easily integrated and gives       
the following expression for $\phi_1^{(1)}$,
\beq
\phi_1 \, = \, L^{2k + n -4}_* {A^{n-1} \over a^{k-1} } \frac{\theta(r+t)}{(t-r)^{n+k-3}r}
\, =\, L^{2k + n -4}_* {A^{n-2} \over a^{k-1} } \frac{\theta(r+t)}{(t-r)^{n+k-3}} \, \left ( {A\over r} \right ) 
\eeq
In the very last expression, we have singled out a factor $A/r$ for convenience of comparing
 with $\phi_0$ which is of order $A/r$.  
   Taking $2k = n$  and comparing this with $\phi_0$, keeping in mind that 
  $A^2/a \sim \sqrt{s}$, we get that $r_* =  \hat{r}$ given by (\ref{rstargood}).
   
 For $k > n/2$ we can apply the same counting, but one has to make sure 
 that the background is ghost-free, on case by case basis, and we won't enter in the details here. 
 Some interesting ideas  appeared recently about  self-protection of such systems \cite{stefan}
 as well as about generalized high-derivative candidates for classicalizing theories\cite{cedric}.

 For us important thing is that,  since,  $\hat{r} \geq r_*$,  confronting with (\ref{rstaralpha}), we see,  that  for 
scalar  theories  $\alpha \, \leq \,  (k-1)/(n+k-3)$ 
 and since $n > 2$,  we have $\alpha < 1$.

  The reason why spin-2 case avoids this power-counting is clear from the equation  
  (\ref{gravitoneq}),  which is a tensor equation, and derivatives acting on 
  wave-packet  on the r.h.s. are not  Lorentz-contracted with each other. 
    As a result,   the r.h.s.  of the equation, 
   \begin{equation}
  \square  \, h_{00}^{(1)} \, = \, - \, L_P \, (T_{00} \, - \, {1 \over 2} 
  \, T_{\alpha}^{\alpha} \, ) \, ,
   \label{gravitoneqzerozero}
  \end{equation}
 when evaluated on a similar free wave-packet  (of energy $\sqrt{s} \, = \, 1/a$ and small occupation number), 
 is of order $\sim 1/(ar)^2$, as opposed to  $1/ar^3$  that it would be had we replaced 
 graviton with a scalar with a two-derivative cubic interaction.  The resulting 
 $r_*$ therefore is $r_* \, = \, L_*^2 \sqrt{s}$.   
 
 In other words, we observe,  that graviton is a most efficient classicalizer!
 
  \subsection{Generalized Holographic Bound on Information Storage}
  
  Given the fact, that  classicalizing theories are based on the  concept  of a  minimal length ($L_*$) and a  classical 
radius ($r_*(s)$)  associated with the energy,  they allow for a generalization of  the black hole bound on information storage. Because the latter  bound involves the area, often it is refereed to as the {\it holographic}  bound.    We shall see, that in  a classicalizing theory the analog of this
bound can be derived and can be interpreted  as the bound on the 
number of soft quanta $N$ that is required to make up a given classicalon configuration.   

   Because of this, the landscape  of classicalizing theories can also be parameterized according to this bound. 
 As we shall see,  not surprisingly, gravity gives a most stringent bound. 
   Of course, in each case, the bound has to be understood  as the bound on amount of 
   information that can be stored in the quanta that exhibit classicalization.
   
     In order to derive the bound, let us consider a classicalizing theory with 
   $r_*$ given by (\ref{rstaralpha}).   Imagine localizing a bit of information within a region of 
   size $r$.   Since information is encoded in particles, such a localization cost at least energy 
   $\sim 1/r$.  Correspondingly,  for $N$ bits this energy is at least $ \sqrt{s} \sim \, N/r$.
   This lower bound on needed energy is independent of how the particles carrying this information interact. 
     In a weakly-coupled non-classicalizing theory,  one can store an {\it arbitrary}  amount of information within a fixed region $r$, 
by pumping-in a sufficient amount of energy. 
     However,  in a classicalizing theory this is not possible, since required energy  makes the 
   $r_*$  radius of the information-storage grow according to (\ref{rstaralpha}), and thus, resists localization.  The $r_*$
 corresponding  to the above-mentioned energy-storage is
 \begin{equation}
 r_*(N) = L_*(L_*N/r)^{\alpha} \, .  
 \label{rstarN}  
\end{equation}
 A given amount of  information ($N$-bits)  can be contained  within a volume $r$ as long as $r < r_*(N)$. 
 Thus,  the maximal amount of information stored within the volume $r$ is 
 \begin{equation}
N \,  = \, (r/L_*)^{1 + {1\over \alpha}} \, .  
 \label{Nbound}  
\end{equation}
This information hits its lower bound for gravity case,  $\alpha =1$, and is unbounded for weakly-coupled 
non-classicalizing theories, that correspond to $\alpha = 0$. 
   More efficient is a classicalizer, more restrictive is the information storage within a given volume, 
   and gravity provides an absolute bound. 
 
   Finally, notice that from (\ref{classrate}) using (\ref{rstaralpha}) we can express the decay rate 
of any classicalon into few particles as   
    \begin{equation}
        \Gamma_{{\rm classicalon} \rightarrow {\rm few}} \sim {\rm e}^{- \, (r_*/L_*)^{1 + {1\over \alpha}} } \, .
        \label{classrategen}
        \end{equation}
 From here it is clear that  among all classicalons,  per fixed size, the black holes are the ones that decay into few particle states in a least suppressed way.   This is because,  among all the classicalons of a  given size, the black holes are the ones that are made of the least number 
 of soft quanta.   This also explains,  why for a weakly-coupled non-classicalizing theory  ($\alpha\,  = \, 0$) the above suppression is formally infinite.  Because in such  theories  classicalons  are never formed (even for energies corresponding to  infinite $N$).

  \subsection{Black Hole Entropy as Classicalon Entropy?} 
  
  In a quantum field theory framework  the simplest way to assign an entropy to a black hole is to compute first the 
Hawking radiance temperature $T_H$ and to define the entropy by standard Clausius's rule:
$T_{H}=\frac{\partial M}{\partial S_{bh}}$. This procedure leads to the well known black hole entropy 
$S_{bh} = \frac{A}{4}$ for $A$ being the horizon area in Planck length units. This relation between entropy 
and horizon area fits nicely with the black hole area theorems. The conceptual puzzle underlying the notion 
of black hole entropy is to understand its statistical meaning. Is this entropy the log of the number of internal 
black hole states associated with the same values of global mass, charge and angular momentum? Is the black hole 
entropy the log of the different ways we can create a given black hole? Is the entropy the log of the number of 
holographic quantum states we can fit on the horizon? Is it an  entanglement entropy for the interior region? 
It is most likely,  that all these questions can lead us to complementary understandings of 
the same fundamental issue.   

 On the other hand,  as we tried to argue, the black hole formation can be viewed as a classicalization process, 
and  black hole entropy has a precursor in classicalon language 
 in terms of number of soft quanta produced in classicalization process.     This fact may give us a possibility 
of pushing this connection further and understanding the origin of black hole 
 entropy (at least qualitatively) in terms of classicalon states.

In other words,  the notion of  a classical interaction range $r_*$ leads to an alternative approach to black 
hole entropy and provides at  least a qualitative way to define it in purely microscopic terms. The recipe is the following.  
 Interpreting a black hole of mass $M$ as a non-perturbative state in the spectrum of pure Einsteinian gravity the associated
 entropy will be defined as the log of the number of quantum states of the set of the interacting quanta of 
wave-length $\sim r_*(M)$ that  we can produce at energy $M$.  The physical meaning of singling out such 
wavelengths is that this are dominant contributors in the classicalization process that takes place at distance $r_*(M)$, 
 which defines the interaction range.  
Correspondingly,  the number of these quanta is determined as,
\begin{equation}
N(M)= M r_*(M) \, .
\end{equation}
Therefore,  the dimension of the corresponding Hilbert space of states will be $d= \xi^N$, for $\xi$ being a number of 
order one determined by the number of states of constituent quanta, such as e.g., helicities of the quanta of the theory.
 This leads to the entropy $S(M)= N(M) {\rm ln}(\xi)$. Hence we can estimate the black hole entropy once we know the 
classicalization radius of gravity. As shown in this paper the interaction range $r_*(M)$ for  gravity is given by the 
Schwarzschild gravitational radius of $M$ and the previous recipe leads to a qualitatively-correct entropy counting. 
Note, that this approach to black hole entropy is purely statistical and does not use any notion of horizon or other 
geometric entities. The Clausius rule relating entropy and temperature appears as a consequence of the exponential 
suppression for the decay process of a set of $N$ "soft" quanta into a few  "hard" ones. The rough estimate of this amplitude is
\begin{equation}
(\frac{1}{ML_P})^{N(M)} \, , 
\end{equation}
which agrees with the exponential suppression of  (\ref{suppression}). 

The key point of the above  approach to black hole entropy is,  that it generalizes to any quantum field theory. 
Namely we can define an entropy function $S(M)$ depending on energy as $Mr_*(M)$. Generically for renormalizable 
theories $r_*(M) \sim\frac{g}{M}$, where $g$ is  some weak dimensionless coupling $g \, \ll \, 1$,  and
 consequently $S(M)\sim 1$, i.e. no real notion of entropy emerges in such a case. Contrary, for theories 
that classicalize,  i.e., for those with $r_*(M)$ growing as function of  $M$ for $M\, \gg \, L_*^{-1}$, 
entropy starts to develop  once we cross above the naive unitarity bound $M_*$. Note also,  that this 
generalized notion of entropy for theories that classicalize always grows with energy.

\section{Conclusions}

 The purpose of this paper is to establish  close parallel between  classicalization processes 
 in gravity and in non-gravitational theories.  For this, we have deconstructed the process 
 of  $s$-wave high energy scattering in gravity,  which by all accounts is expecting to end up by black hole 
 formation.   For us,  however, most important  aspect was understanding of the  pre-horizon-formation stage of 
this  process from the point of view of classicalization.  
 We have seen,  that this process is nothing but  a version of  classicalization process that also takes place 
in analogous scattering of derivatively-coupled  Nambu-Goldstone-type scalars. 
 
    The main finding of this analysis is, that  $r_*(s)$ for  gravity agrees with the gravitational radius
    of energy $\sqrt{s}$.  This allows us to understand most of the black hole unitarization properties as 
direct consequences of classicalization.
   
  In particular we see,  that there is a clear precursor of entropy in terms of number of 
  soft quanta produced in the classicalization process.  This notion is universally applicable 
  to generic classicalizing theories, regardless of the existence of the horizon, and follows 
  from classicality of the configuration produced.  
   
     This universality allows for a parameterization of classicalization landscape in terms 
     of a parameter that measures the  classicalization-efficiency, and for  generalization 
    of  black hole properties, such as holographic bound on information storage, to non-gravitational situations.   
Being a most efficient classicalizer,   gravity is the limiting case on this landscape,  and correspondingly, 
 gives a most stringent bound on information storage.  
  
\newpage
\centerline{\bf Acknowledgments}
We thank  Gian Giudice for ongoing discussions and collaboration on classicalization. 
We thank Thibault Damour, Andrei Gruzinov,  Cristiano Germani and  Goran Senjanovic for discussions.
The work of G.D. and C.G.  was supported in part by Humboldt Foundation under Alexander von Humboldt Professorship,  
by European Commission  under 
the ERC advanced grant 226371,  and  by the NSF grant PHY-0758032. 
The work of C.G. was supported in part by Grants: FPA 2009-07908, CPAN (CSD2007-00042) and HEPHACOS-S2009/ESP1473.
The work of A.K is supported by a PEVE-NTUA-2009 grand.

\end{document}